\documentclass[doublecol,figures]{epl2}
\usepackage{amsfonts}
\usepackage{amsmath}
\usepackage{amssymb}

\institute{
\inst{1} Department of Physics and Michigan Center for Theoretical Physics -
University of Michigan, 450 Church Str., Ann Arbor, Michigan 48109 USA
}
\abstract{In recent years there have been a number of proposals to utilize the
specificity of DNA based interactions for potential applications in
nanoscience.  \ One interesting direction is the
self-assembly of micro- and nanoparticle clusters using DNA scaffolds. \
In this letter we consider a DNA scaffold method to self-assemble clusters of
"colored" particles. \ Stable clusters of identical microspheres have recently been
produced by an entirely different method. \ Our DNA based
approach self-assembles clusters with additional degrees of freedom
associated with particle permutation. \ We demonstrate that in the
non-equilibrium regime of irreversible binding the self-assembly process is
experimentally feasible. \ These color degrees of freedom may allow for more
diverse intercluster interactions essential for hierarchical self-assembly
of larger structures. \ 
}

\begin{document}

\title{How to build nanoblocks using DNA scaffolds}
\author{Nicholas A. Licata\inst{1} and Alexei V. Tkachenko\inst{1}}
\maketitle

DNA has attracted significant attention for its potential applications in 
nanoscience (\cite{natreview},\cite{blocks},\cite{micelle},\cite%
{nanocrystals},\cite{designcrystals},\cite{template},\cite{rational},\cite%
{storhoff},\cite{supra},\cite{biotech}). \ One recent non-DNA based advance is the self-assembly 
of stable clusters composed of identical microspheres \cite{packing}. \ In this letter we consider 
the self-assembly of micro- and nanoparticle 
clusters similar to those of \cite{packing}, where DNA scaffolds govern the self-assembly process. \ 
The plan for the letter is the following. \ We first introduce the basic
strategy of our self-assembly proposal. \ The goal is to maximize the yield
for a particular type of cluster we call the star cluster. \ We analytically
calculate the yield of the star cluster in the regime of irreversible
binding. \ The analytical results are compared to the numerical results for
the full aggregation equations. \ From an experimental perspective, the most
important result is the determination of an optimal concentration ratio for
experiments (see Eq. \ref{popt}). \ To conclude we discuss the experimental
feasibility of the self-assembly proposal. \ 

\begin{figure}[h]
\includegraphics[width=3.365in,height=2.5339in]{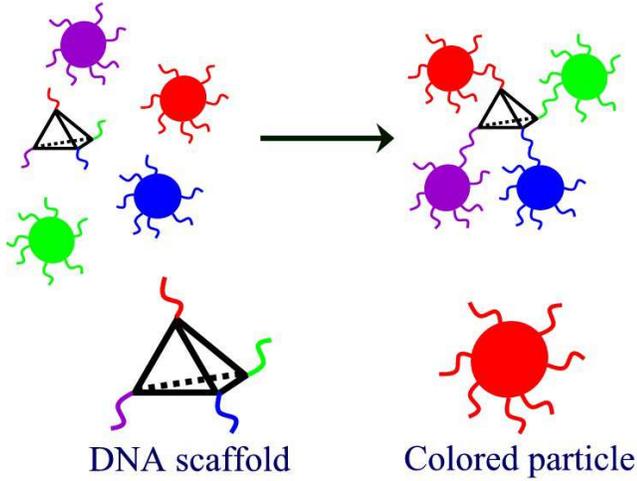}
\caption{ \textbf{DNA scaffolding}. \ A graphical depiction of the scheme
for self-assembling star clusters using DNA scaffolds. \ In the diagram (not
drawn to scale) the scaffold funcionality $f=4$. \ }
\label{dnascaffold}
\end{figure}

The basic idea behind the procedure is as follows (see Fig. \ref{dnascaffold}%
). \ Particles are functionalized with single-stranded DNA (ssDNA) markers
which determine the particle color. \ There may be many DNA attached to each
particle, but on any given particle the marker sequence is identical. \ One
then introduces DNA\ scaffolds to the system. \ The scaffold is a structure
with $f$ ssDNA\ markers, each marker complementary to one of the particle
colors. \ Hybridization of the ssDNA markers on the particles to those on
the scaffold results in the formation of colored particle clusters. \
Because there are many DNA\ attached to each particle, clusters can form
which contain more than one scaffold. \ The essential goal of the procedure
is to maximize the concentration of a particular type of cluster which we
denote the star cluster. \ The star cluster contains one and only one
scaffold to which $f$ particles are attached, each particle having a
distinct color. \ 

We should note that the role of the scaffold could also be played by a
patchy particle (\cite{patchy},\cite{patchy2},\cite{patchy3}). \ For
example, these patches are regions on the particle surface where one can
graft ssDNA markers. \ In this case there may be several DNA connections
between a patch and colored particle. \ Our conclusions will still be valid,
provided the patch size is chosen so that a patch interacts with at most one
particle. \ 

Previously we performed an equilibrium calculation to determine the yield of
the star cluster\cite{nanoclusters}. \ The results of that study indicated
that the concentration of scaffolds must be kept very small to prevent the
aggregation of larger clusters. \ From an experimental perspective this
result is somewhat disappointing, since the overall yield of the star
cluster is proportional to the scaffold concentration. \ The situation is
considerably improved in the regime of irreversible binding of particles to
scaffolds. \ In what follows we present a calculation for the yield of the
star cluster far from equilibrium. \ 

To understand the basic physics behind the aggregation process we consider
the mobility mismatch between the particles and the scaffolds. \ In
solution, a particle with radius $R\simeq 1\mu m$ has a diffusion
coefficient given by the Stokes-Einstein relation $D=k_{B}T/6\pi \eta R$. \
On the other hand, the size of the scaffold $a\simeq 10nm$. \ As a result
the scaffolds diffuse $R/a\simeq 100$ times faster than the particles. \ To
first approximation the resulting aggregation is a two stage process. \ In
the first stage the particles recruit different numbers of scaffolds via the
fast scaffold diffusion and subsequent DNA\ hybridization. \ Since we
consider the regime of strong binding where these bonds are irreversible,
the result is a Poisson distribution over the concentration of particles
with $m$ scaffolds attached. \ Let $C_{i}$ denote the concentration of
particles with color $i$, and $c$ denote the total concentration of
scaffolds. \ The total particle concentration $C_{tot}=\sum_{i=1}^{f}C_{i}$.
\ The concentration $C_{i}^{(m)}$ of particles of color $i$ with $m$
scaffolds attached is%
\begin{eqnarray}
C_{i}^{(m)} &=&C_{i}\frac{p^{m}\exp (-p)}{m!}  \label{pdef} \\
p &=&\frac{c}{C_{tot}}
\end{eqnarray}

In the second stage there are no free scaffolds left in solution, and these
particles decorated with scaffolds aggregate to form the final clusters. \
The seed to build a star cluster is a particle of any color with exactly one
scaffold attached. \ This seed must aggregate with $f-1$ particles of
different colors, each of which has no scaffolds. \ We now calculate the
concentration of the star cluster $C_{\ast }$. \ The yield of the desired
star cluster is quantified in terms of the star mass fraction $M_{\ast
}=(fC_{\ast })/C_{tot}$. \ 
\begin{equation}
M_{\ast }=\frac{f}{C_{tot}}\sum\limits_{i=1}^{f}C_{i}^{(1)}\prod\limits 
_{\substack{ j=1  \\ j\neq i}}^{f}\frac{C_{j}^{(0)}}{C_{j}}=x\exp (-x)
\label{cstar}
\end{equation}%
Here $x=fp$ is the scaffold functionality $f$ multiplied by the
concentration ratio $p$. \ By choosing $p=1/f$ the mass fraction attains a
maximum of $\exp (-1)\simeq 0.37!$ \ This result indicates that by selecting
the appropriate scaffold concentration, in the nonequilibrium regime up to $%
37\%$ of the particles will aggregate to form star clusters. \ This is a
significant improvement over the situation in the equilibrium regime. \ 

This treatment of the problem captures the physics of star cluster
formation, but it does not account for the loss of star clusters due to
aggregation. \ In particular, as long as there are scaffolds with markers
available for hybridization, when these scaffolds encounter a star cluster
they can aggregate to form a larger cluster. \ We now estimate how this
aggregation effects the final concentration of star clusters. \ 

Consider the beginning of the second stage in our aggregation process. \
There are no longer any free scaffolds in solution, but a scaffold can have
up to $f-1$ DNA markers still available for hybridization. \ We would like
to determine how the star cluster mass fraction $M_{\ast }(y)$ changes as a
function of the fraction of saturated scaffolds $y$. \ Here a saturated
scaffold has particles hybridized to all $f$ of its DNA markers, and is
therefore unreactive. \ If $s$ is the expectation that a slot on the
scaffold is filled, then the fraction of saturated scaffolds is $y=s^{f-1}$.
\ The average number of open slots on a scaffold is $(f-1)(1-s)$. \ Consider
filling an open slot on the scaffold. \ The probability that the particle
which filled the slot was part of a star cluster is $M_{\ast }(y)$. \ The
average rate $r(y)$ at which star clusters are lost to aggregation is then%
\begin{equation}
r(y)=-M_{\ast }(y)\frac{d}{dy}\left[ (f-1)(1-s)\right] =M_{\ast
}(y)y^{-\alpha }\text{.}
\end{equation}%
Here the exponent $\alpha =(f-2)/(f-1)$. \ We can then construct a
differential equation for $M_{\ast }$ taking into account this loss due to
aggregation. \ 
\begin{equation}
\frac{dM_{\ast }}{dy}=\frac{dM_{\ast }^{(o)}}{dy}-xr(y)
\end{equation}%
In the absence of this loss term the result of the calculation should
recover our previous result Eq. \ref{cstar}. \ This zeroth order
approximation is just $M_{\ast }^{(o)}(y)=xy\exp (-xy)$ which gives the
correct star cluster concentration once all of the scaffolds are saturated ($%
y=1$). \ To simplify the analysis a bit we take $\alpha =1$ which is an
excellent approximation in the limit of large scaffold functionality $f$. \
This is an inhomogeneous first order differential equation which can be
solved by introducing an integrating factor $u(y)=y^{x}$. $\ $The initial
condition which must be satisfied is $M_{\ast }(0)=0$. \ We are interested
in the final star mass fraction $M_{\ast }$, which is $M_{\ast }(y=1)$. \
The result is%
\begin{eqnarray}
M_{\ast } &=&x\sum_{k=0}^{\infty }\frac{(-x)^{k}}{k!}\left[ \frac{1}{x+k+1}-%
\frac{x}{x+k+2}\right]  \label{massfractheory} \\
&=&x\exp (-x)+x^{2}E_{-x}(x)-x^{1-x}\Gamma (1+x)  \notag
\end{eqnarray}%
Here $\Gamma (x)$ is the gamma function and $E_{\nu
}(x)=\int\limits_{1}^{\infty }t^{-\nu }\exp (-xt)dt$ is the exponential
integral of order $\nu $. \ 

We can perform a similar type of analysis in the case when there is only one
particle color. \ In this case the $f$ ssDNA markers on the scaffold all
have identical sequences complementary to this color. \ It turns out that
the result for the mass fraction is the same. \ Because the mass fraction is
the same in both cases, we can gain insight into the behavior of the system
with many colors by analyzing the much simpler one color system. \ To test
our predictions, we numerically solved a system of differential equations
which models the irreversible aggregation between particles (one color) and
scaffolds. \ 
\begin{equation}
\frac{dC_{IJ}}{dt}=\frac{1}{2}\sum\limits_{\substack{ i+i^{\prime }=I  \\ %
j+j^{\prime }=J}}K_{iji^{\prime }j^{\prime }}C_{ij}C_{i^{\prime }j^{\prime
}}-C_{IJ}\sum\limits_{i,j}K_{ijIJ}C_{ij}
\end{equation}

This equation is the Smoluchowski coagulation equation\cite{Smoluchowski}
adapted to our system. \ $C_{ij}$ is the concentration of the cluster with $%
i $ scaffolds and $j$ particles. $\ K_{iji^{\prime }j^{\prime }}$ is the
rate constant for the irreversible reaction $C_{ij}+C_{i^{\prime }j^{\prime
}}\rightarrow C_{i+i^{\prime }j+j^{\prime }}$. \ We assume that the rates
are diffusion limited in which case we can estimate the rate for any pair of
clusters by $K_{iji^{\prime }j^{\prime }}=4\pi D_{s}R_{l}$. \ The larger
cluster with hydrodynamic radius $R_{l}\sim n_{l}^{1/3}$ plays the role of a
sink. \ Here $n_{l}$ is the number of particles in the larger cluster and $%
D_{s}=k_{B}T/6\pi \eta R_{s}$ is the diffusion constant for the smaller
cluster. \ To simply matters we only consider tree like structures, i.e. we
do not consider the formation of clusters with internal loops. \ We have
truncated the set of equations by considering clusters with a maximum of $10$
scaffolds. \ 

\begin{figure}[h]
\includegraphics[width=3.3446in,height=2.5173in]{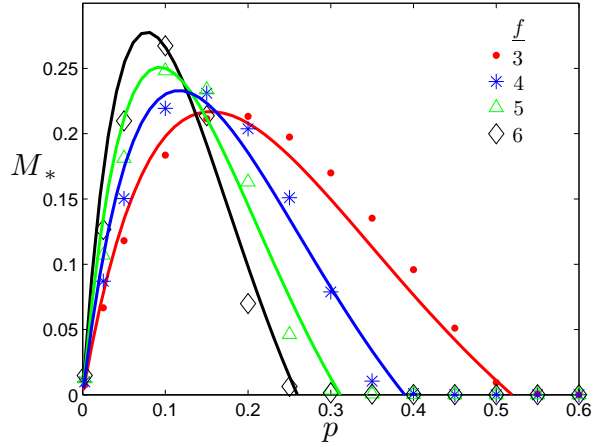}
\caption{ \textbf{Star cluster mass fraction}. \ The mass fraction $M_{\ast }
$ as a function of $p$ for scaffolds with functionality $f=3$ (red), $4$
(blue), $5$ (green), and $6$ (black). \ The results determined numerically
from the full solution of the Smoluchowski coagulation equation (markers)
can be compared to results of the anlaytical calculation (lines) Eq. \protect
\ref{massfractheory}. \ The resulting agreement is good up to an overall
normalization factor $\protect\gamma _{f}$ in the range $1.2$ to $1.5$ which
normalizes the analytical curves. \ }
\label{massfracstar}
\end{figure}

By solving these equations we can determine the concentration of stars $%
C_{\ast }=C_{1f}$ in this notation and test the validity of our two stage
ansatz. \ As indicated in Fig. \ref{massfracstar}, the result of our
analytical calculation matches the results of the full numerical calculation
up to an overall normalization factor of order unity. \ Several points are
in order. \ 

The optimal concentration ratio $p$ for experiments is easily determined
from $\frac{dM_{\ast }}{dx}=0$. \ The result is $x_{\max }\simeq 0.47$. \
For scaffolds of functionality $f$ the concentration ratio should be chosen
as:%
\begin{equation}
p=\frac{0.47}{f}\text{.}  \label{popt}
\end{equation}
Note that the maximum attainable star cluster yield $M_{\ast }(x_{\max
})\simeq 1/4$ does not decrease with increasing $f$. \ In fact, the
numerical results predict a slight increase in star cluster yield for larger 
$f$. \ Solving the aggregation equations becomes computationally expensive,
but it can still be done by reducing the maximum number of scaffolds in a
cluster. \ For example, considering clusters with up to $5$ scaffolds for $%
f=10$ gives $M_{\ast }(x_{\max })\simeq 0.3$. \ These results are important
from the perspective of experimental feasibility for the self-assembly
method. \ This is to be contrasted with the earlier equilibrium treatment. \
There the condition to suppress the aggregation of larger clusters imposed a
fairly strict constraint\cite{nanoclusters} on the concentration ratio $%
p\lesssim f^{1/2}\left( \frac{2}{f}\right) ^{f-1}$. \ From the perspective
of self-assembling stars with large $f$ this renders the regime of
irreversible binding far more appealing than the equilibrium regime. \ 

If an experiment is performed with the optimal concentration ratio, the
clusters which self-assemble are easily separated by density gradient
centrifugation\cite{centrifugation}. \ In this regime most of the particles
are monomers, in star clusters, or in saturated two scaffold clusters. \
These clusters contain, $1$, $f$, and $2f-1$ particles respectively. \ The
disparity in hydrodynamic radius and sedimentation velocity of these
clusters makes the separation procedure experimentally feasible. \ 

In this letter we considered a DNA\ scaffold method for self-assembling star
clusters of $f$ colored particles. \ By taking advantage of the mobility
mismatch between particles and scaffolds, we were able to formulate a
nonequilibrium calculation of the star mass fraction. \ The results of the
calculation were compared to the numerical results of the full Smoluchowski
coagulation equation for the system. \ Good agreement is established between
the analytical calculation and the numerics. \ In the regime of irreversible
binding the yield of the desired star cluster is drastically improved in
comparison to earlier equilibrium estimates. \ In nonequilibrium we find an
experimentally feasible regime for the self-assembly of star clusters with a
maximum mass fraction $\simeq 1/4$. \ We determined the optimal
concentration ratio for an experimental implementation of our proposal. \
The additional color degrees of freedom associated with particle permutation
in these clusters makes them ideal candidates as building blocks in a future
hierarchical self-assembly scheme. \ In addition, these clusters can serve
as the starting point to self-assemble structures of arbitrary geometry\cite%
{licata}. \ The experimental realization of self-assembling star clusters
using DNA\ scaffolds would constitute an important step towards realizing
the full potential of DNA\ mediated interactions in nanoscience. \ 

This work was supported by the ACS\ Petroleum Research Fund (Grant PRF No.
44181-AC10). \ We acknowledge L. Sander, G. Ghoshal, and B. Karrer for
valuable discussions. \

\end{document}